\documentclass[fleqn,10pt]{wlscirep}
\usepackage[utf8]{inputenc}
\usepackage[T1]{fontenc}
\usepackage{lineno}
\usepackage{booktabs}
\usepackage{graphicx}
\usepackage{xcolor}

\title{People are poorly equipped to detect AI-powered voice clones}

\author[1]{Sarah Barrington}
\author[2,3]{Emily A. Cooper}
\author[1,4,*]{Hany Farid}
\affil[1]{School of Information, University of California, Berkeley, Berkeley, CA, 94720}
\affil[4]{Electrical Engineering and Computer Sciences, University of California, Berkeley, Berkeley, CA, 94720}
\affil[2]{Herbert Wertheim School of Optometry, University of California, Berkeley, Berkeley, CA 94720}
\affil[3]{Helen Wills Neuroscience Institute, University of California, Berkeley, Berkeley, CA 94720}

\affil[*]{corresponding author: hfarid@berkeley.edu}

\begin{abstract}
As generative artificial intelligence (AI) continues its ballistic trajectory, everything from text to audio, image, and video generation continues to improve at mimicking human-generated content. Through a series of perceptual studies, we report on the realism of AI-generated voices in terms of identity matching and naturalness. We find human participants cannot consistently identify recordings of AI-generated voices. Specifically, participants perceived the identity of an AI-voice to be the same as its real counterpart approximately $80\%$ of the time, and correctly identified a voice as AI generated only about $60\%$ of the time.
\end{abstract}
\begin{document}

\flushbottom
\maketitle

\thispagestyle{empty}

In January 2024, in the lead up to the November United States presidential election, an estimated tens of thousands of Democratic party voters received a robocall in the voice of President Biden instructing them not to vote in the upcoming New Hampshire primaries. The voice was AI-generated.

The perpetrators of this attempted election interference were Steven Kramer (a political consultant), Paul Carpenter (a magician and hypnotist who was paid \$150 to create the fake audio), and a telecommunications company called Lingo Telecom~\cite{npr-biden-robocall}. Carpenter used ElevenLabs, a platform offering instant voice cloning for as little as \$5 a month. Kramer was fined \$6 million and subsequently charged with two dozen crimes including impersonating a candidate and voter suppression, while the telecommunications company, Lingo Telecom, received a \$1 million fine for transmitting the calls. This is just one of many examples of how the rise of generative AI is being weaponized, from election interference, to disinformation campaigns~\cite{nbc-harris-story}, to small-\cite{aarp2024chatbots} and large-scale~\cite{cnn-hk-story} financial fraud.

There is large literature on technologies that can automatically determine whether media -- such as audio, video, and images -- has been manipulated either by humans or generative AI~\cite{farid2022creating}. These techniques, however, largely operate asynchronously and not as an audio or video call is unfolding in real time. The synchronous detection of fraudulent media, such as the phone calls that attempted to suppress voter turnout in New Hampshire, still poses significant technological challenges. Until technology can monitor every landline, mobile device, and video call (which itself would raise additional privacy concerns), people are largely left to their own defenses to sort out the real from the fake.

The question then naturally arises: how well-equipped are people for the perceptual challenge of distinguishing real from AI-generated content? The answer, of course, depends on both the quality of the fake and the modality of the media. For example, studies focusing on visual perception of images of people have concluded that participants are at chance at distinguishing real and AI-generated head shots~\cite{aisynthesizedfaces,bray2023testing}. Results for video (with audio) are more mixed -- likely due to differences in the types of videos that have been assessed. While some studies report that performance is only slightly better than chance for videos of people talking~\cite{deepfakedetectionhumancrowd,kobis_fooled_2021}, a recent large-scale study investigating how well people could distinguish fabricated political speeches from real ones report a consistent accuracy of $80\%$ and above~\cite{groh2024human}. 

Here, we focus on people's ability to distinguish real voices from AI-generated voice clones, as would be required to detect a fraudulent phone call or voicemail. Interestingly, prior work suggests that people are better at this task than detection of AI-manipulated images, but can still often be tricked. For example, Mai et al.~\cite{mai} report that human participants were able to accurately distinguish real voices from AI-generated voice clones with an accuracy of $70.4\%$. This study, however, only used a single English and a single Chinese speaker identity, and the spoken phrases consisted of a single sentence ranging in length from $2$ to $11$ seconds (by comparison the fake Biden robocall was $40$ seconds in length). Müller et al.~\cite{muller} report a similar accuracy of $80\%$. This second study has the advantage that it employed multiple speaker identities ($107$), but the spoken phrases were still relatively short at one to two sentences in length. Mostly consistent with these earlier studies, a recent study by Warren et al.~\cite{traynor2024} found that human participants detect AI-generated voices with an accuracy of $73\%$. For each of these studies, the AI-generated voices were not created using state-of-the-art, commercially available techniques, and both studies focused on naturalness (is the voice real or not) and did not examine identity perception (who is speaking). 

Groh et al.~\cite{groh2024human} also investigated audio-only performance with political speeches. Performance when people were only given the isolated audio of the speeches was worse than with the full video with audio, but still above chance. Political speeches, however, are atypical given that the speakers are highly familiar to most listeners, and that political speeches have characteristic cadences and content that do not necessarily reflect natural conversational speech.

We expand on these previous studies by employing the state-of-the-art voice cloning of ElevenLabs (used in the Biden robocall), increasing the number of speakers to over $200$, and considering how different tasks (identity and naturalness) impact our ability to distinguish AI-generated voices. This study reveals that, generally speaking, people are poorly equipped to identify AI-generated voice clones, both in terms of identity matching and naturalness.

\section*{Methods}

\subsection*{Real Speaker Dataset}

The stimuli for our study were sampled from the DeepSpeak dataset, which comprises recordings of $220$ unique speakers collected through the Prolific research recruitment platform~\cite{2024deepspeak}. More details about this dataset can be found in the accompanying manuscript, but here we provide a brief summary. 

All speakers were native English speakers and U.S. residents. Their ages ranged from 18-75 years (mean=38, sd=11.4), with 109 identifying as male, 107 female, and 4 non-binary. Racial identities included 158 White/Caucasian, 39 Black/African American, 26 Asian, 4 American Indian/Alaska Native, 2 Native Hawaiian/Other Pacific Islander, and 5 other. 

Each speaker was instructed to record themselves responding to $32$ prompts. The prompts were divided into four categories: (1) standardized scripted responses in which each speaker read the same prompt extracted from transcripts of the TIMIT dataset~\cite{timit}; (2) randomized scripted responses in which each speaker read a randomized prompt from TIMIT; (3) unscripted responses in which each speaker responded to four open-ended questions, and asked for a response that was close to $30$ seconds in length; and (4) combined responses consisting of four open-ended unscripted questions in which each speaker read out loud a question and then answered the question. 
 
Both audio and video were recorded using a custom-built web application. The audio/video recordings were converted from their initial .webm format to .mp4 at a bitrate of $192$ kbps from which the audio was extracted as a .wav file. All real audio files were converted to a .mp3 format with a sample rate of 44kHz, with an amplitude normalized between $-1$ and $1$, and with silences at the start and end removed. In this study, we only used a subset of the audio clips, which had a mean duration of 5.37s (min: 0.88s, max: 52.57).

\subsection*{Voice Cloning}
\label{sec:voice-cloning}

A voice clone of each of the $220$ speakers was generated using the ElevenLabs' {\em Instant Voice Cloning} API. For each speaker, a cloned voice model was first synthesized using the audio from just the first two prompts in the DeepSpeak dataset as input. Transcripts of speakers' responses to other prompts were then used to create a cloned version of each original audio clip ($32$ total per speaker). For scripted responses, we assumed that the speaker correctly repeated the prompt; for unscripted responses, OpenAI's {\em Whisper} \cite{whisper} was used to transcribe the audio. Cloned speaker audio files were converted to the same format, sample rate, and amplitude as the real speaker audios.

\subsection*{Voice Matching}
\label{sec:voice-matching}

For one of our studies, participants were presented with two real voices with different speaker identities. When people heard two speakers with different identities, we wanted the voices to be similar to each other. Thus, for each speaker in our dataset, we also determined another speaker with a perceptually similar voice. This voice matching was performed by first extracting a $192$-D TitaNet embedding~\cite{koluguri2022titanet} of the same scripted sample. The closest matching voice was determined by finding the voice of another speaker (with replacement) with the maximal cosine similarity between extracted embeddings (the mean similarity was $0.6$ in the range $[-1,1]$). Audios were paired by scriptedness to ensure comparisons were made between similar contexts.

\subsection*{Study Design}
\label{sec:study-design}

We examined people's perception of AI-powered voice clones in two perceptual studies. 

In the {\em identity} study (Figure~\ref{fig:methodsandresults}(a), left), participants listened to two voices back-to-back (saying something different) and were asked to specify if the voices were from the same identity. Participants were randomly assigned to receive one of $10$ batches comprising a randomized set of $44$ stimuli (i.e.,~$44$ unique random voice pairings). Thirty of these stimuli contained scripted single-sentence responses, $10$ were unscripted responses, and $4$ were attention checks (see below). There was no stimulus overlap between batches. Because the voice pairings were randomized, each stimulus could fall into six possible conditions. The three conditions of core interest for our analyses were: 
\begin{itemize}
\setlength\itemsep{-0.5em}
\item [$\bullet$] [same identity]~/~[real speaker] ($A-A$),
\item [$\bullet$] [same identity]~/~[real and AI-clone speaker] ($A-\hat{A}$), 
\item [$\bullet$] [different identity]~/~[real speaker] ($A-B$). 
\end{itemize}
There were also trials that fell into three additional conditions, which ensured that the structure of the study could not be learned by participants over time: 
\begin{itemize}
\setlength\itemsep{-0.5em}
\item [$\bullet$] [same identity]~/~[AI-clone speaker] ($\hat{A}-\hat{A}$), 
\item [$\bullet$] [different identity]~/~[AI-clone speaker] ($\hat{A}-\hat{B}$), 
\item [$\bullet$] [different identity]~/~[real and AI-clone speaker] ($A-\hat{B}$). 
\end{itemize}
Participants were asked only to judge identity and were not told that some voices may be AI generated.

In the {\em naturalness} study, participants listened to one voice at a time and were asked to classify it as real or AI generated (Figure~\ref{fig:methodsandresults}(b), left). The randomization and assignment of stimuli was identical to the identity study, except in this case each stimulus was one audio clip rather than two. Half of the audio clips were real and half were AI generated. Participants were not told of this distribution.

For both studies, participants did not receive any explicit instructions to use headphones, earphones or speakers. The full instructions are included in the Supplementary Information.

\subsection*{Listener Participants}

A total of $604$ participants were recruited from the Prolific crowd-sourcing platform, split into two groups of $304$ and $300$ for the identity and naturalness studies. Listener ages ranged from $18-77$ years (mean=$35$, sd=$12$), with $293$ male, $294$ female, $11$ non-binary and $6$ not providing their gender. $411$ listeners identified as White/Caucasian, $132$ as Black/African American, $43$ as Asian, $19$ as American Indian/Alaska Native, $4$ as Native Hawaiian/Other Pacific Islander, $30$ as other, and $5$ preferred not to share.

Before beginning the study, participants were given an overview of their task (either in terms of judging identity or naturalness) and tested their hardware. For the naturalness study, participants were given two examples of real voices and two examples of AI-generated voices to set expectations as to the realism of the voices.

To ensure that participants were paying attention, four attention checks were randomly distributed throughout. These checks consisted of audio clips that clearly described the correct answer to be selected. Participants who failed any of these attention checks were removed from subsequent analysis (the totals reported above only include participants who passed all attention checks). Participants could not respond without having listened to the entire audio(s). Participants were paid \$5 for their time.

\subsection*{Statistical Analysis}

For each participant in the {\em identity} study, we calculated the percentage of stimuli for which they responded that the identities were the same, separately for each stimulus condition (($A-A$), ($A-\hat{A}$), ($A-B$), etc). The distribution of responses across conditions deviated substantially from normality, so we adopted non-parametric statistics to examine differences between conditions. We used single sample signed rank tests to examine differences from chance ($50\%$) separately for each condition, then a Friedman test was used to examine whether there were significant differences between conditions. Pairwise follow up tests were conducted via paired sample rank sum tests.

For the {\em naturalness} study, the response distributions were approximately normally distributed, so we used single and paired sample t-tests to examine differences from chance ($50\%$) for stimuli that were real and AI generated, as well as to directly compare responses for these two stimulus types.

For all analyses, we adopted a significance threshold of $p < 0.05$ and applied Bonferroni correction when the same analysis was run on multiple conditions.

Because we used a large dataset of naturalistic stimuli and designed our study to obtain a large listener participant sample, audio clips in this experiment varied in length and each participant only listened to a subset of stimuli (the randomized batches). To examine how these, and other variables, might affect performance, we therefore fit the trial-by-trial responses with mixed-effects logistic regression models. 

For the {\em identity} study, we focused on fitting performance on trials with the same real identity ($A-A$) and those with the same identity but one AI-generated clone ($A-\hat{A}$) -- that is, to keep the model relatively simple we only fitted the data from these two conditions. For the {\em naturalness} study, the model was fitted to all data. Each model also included random effects (intercepts) per participant and per stimulus batch. In addition to the stimulus condition, the fitted models included the following predictors: total audio duration (continuous), scripted vs unscripted (categorical), listener gender (categorical), listener age (continuous), speaker gender (categorical), speaker age (continuous). The combined audio duration of both audios was used. 

For both models, the continuous predictors, age and duration, were scaled using Z-score normalization and robust scaling, respectively. 

\begin{figure}[p!]
    \centering
    \includegraphics[width=0.8\textwidth]{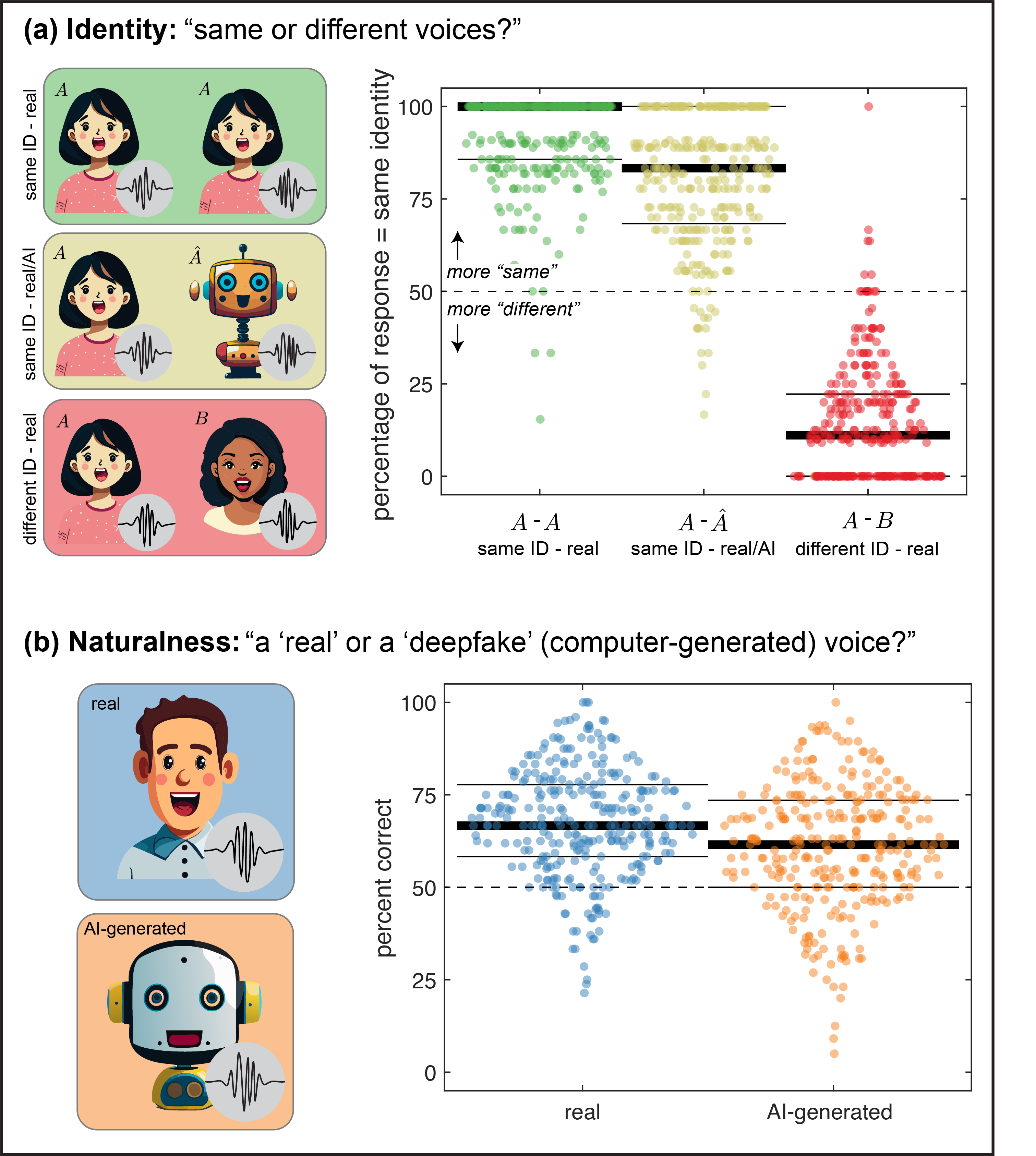}  
    \caption{Study methods and results. a) In the identity study, participants were presented with pairs of audio clips and asked if they were from the same or different people. We present results for three conditions: both voices were the same identity and a real person ($A-A$, green), both voices were the same identity but one clip was an AI-clone ($A-\hat{A}$, yellow), and the voices were from real but different identities ($A-B$). Results are plotted as the percentage of trials on which each participant responded ``same identity.'' Thick lines indicate medians and thin lines indicate $25^{th}$ and $75^{th}$ quantiles. The $50^{th}$ -- $25^{th}$ -- $75^{th}$ quantile of for the three remaining, unplotted conditions are as follows: [same identity]~/~[AI-clone speaker] ($\hat{A}-\hat{A}$) $100.0\%$ -- $100.0\%$ -- $100.0\%$, [different identity]~/~[AI-clone speaker] ($\hat{A}-\hat{B}$) $0.0\%$ -- $0.0\%$ --$20.0\%$, and [different identity]~/~[real and AI-clone speaker] ($A-\hat{B}$) $3.6\%$ -- $0.0\%$ -- $16.7\%$. b) In the naturalness study, participants were presented with audio clips one at a time and asked whether the voice was real or AI generated. Results are plotted in the same manner as panel (a), except now the ordinate indicates percent correct for real (blue) and AI-generated (orange) voices. }
    \label{fig:methodsandresults}
\end{figure}
\newpage

\section*{Results}

\label{sec:headings}
\subsection*{Identity}

When participants listened to a pair of real audio clips with the same speaker ($A-A$), they were on average highly accurate. That is, they overwhelmingly indicated that it was the same person speaking in both clips: $50^{th}$ -- $25^{th}$ -- $75^{th}$ quantile of $100.0\%$ -- $85.7\%$ -- $100.0\%$, as shown in Figure~\ref{fig:methodsandresults}(a), green. Indeed, $58.6\%$ of participants indicated that it was the same identity on every trial for these stimuli. When one of the clips was an AI-clone ($A-\hat{A}$), participants also overwhelmingly judged that it was the same person speaking in both clips: $50^{th}$ -- $25^{th}$ -- $75^{th}$ quantile of $83.3\%$ -- $69.2\%$ -- $100.0\%$ as shown in Figure~\ref{fig:methodsandresults}(a), yellow. However, only $26.6\%$ of participants indicated the same identity on every trial. This finding suggests that AI-clones are highly, but not uniformly, convincing. It is also possible that participants were simply biased to indicate that all speaker pairs were the same. The data, however, show that this was not the case: when participants listened to two different, but similar, voice identities ($A-B$), they rarely indicated that the speakers had the same identity: $50^{th}$ -- $25^{th}$ -- $75^{th}$ quantile of $11.1\%$ -- $0.0\%$ -- $22.2\%$, as shown in Figure~\ref{fig:methodsandresults}(a), red.   

Statistical tests indicate that responses in each of these conditions are significantly different from chance ($A-A$: W = $94.5$, p << $0.001$; $A-\hat{A}$: W = $693.5$, p << $0.001$; $A-B$: W = $301.5$, p << $0.001$). There is also significant differences between conditions (Friedman test = $517.0$, p << $0.001$). Follow up pairwise tests indicate that each pair of conditions differ significantly ($A-A$ vs $A-\hat{A}$: W = $3605.5$, p << $0.001$; $A-A$ vs $A-B$: W = $0.0$, p << $0.001$; $A-\hat{A}$ vs $A-B$: W = $0.0$, p << $0.001$). Responses for the other three conditions are generally consistent with these results and are reported in the caption of Figure~\ref{fig:methodsandresults}(a).

The responses to the $A-A$ and $A-\hat{A}$ conditions were further investigated with a logistic regression model (Table~\ref{tab:mixedlm-results-identity}). As compared to the $A-A$ condition, the $A-\hat{A}$ condition is associated with significantly fewer ``same'' responses. There is also a significant effect of speaker gender: the voices of male and non-binary genders are more often judged ``same'' as compared to female speakers. Interestingly, both combined audio duration and scriptedness have significant effects: longer clips and scripted clips are associated with more frequent ``same'' responses. We will return to explore these effects in more detail in the Discussion.

\begin{table}[t]
    \centering
    \begin{tabular}{lccccc}
        \hline
        \textbf{Predictor} & \textbf{Coef.} & \textbf{z} & \textbf{p-value} & \textbf{95\% CI Lower} & \textbf{95\% CI Upper} \\ 
        \hline
        Intercept                        & $2.836$    & $12.003$   & $<0.001$ & $2.337$   & $3.299$   \\
        Condition Assigned ($A-\hat{A}$)   & $-0.988$   & $-10.814$  & $<0.001$  & $-1.167$  & $-0.809$  \\
        Combined Duration                & $0.304$    & $3.953$   & $<0.001$ & $0.153$   & $0.454$   \\
        Scripted (vs Unscripted)         & $-1.331$   & $-5.831$  & $<0.001$ & $-1.778$  & $-0.883$  \\
        Listener Gender (Male)           & $-0.198$   & $-1.594$  & $0.111$    & $-0.442$  & $0.046$   \\
        Listener Gender (Non-Binary)     & $-0.476$   & $-1.195$  & $0.232$    & $-1.258$  & $0.305$   \\
        Listener Age                     & $-0.006$   & $-1.293$  & $0.196$    & $-0.016$  & $0.003$   \\
        Speaker Gender (Male)            & $0.376$    & $4.227$   & $<0.001$  & $0.202$   & $0.551$   \\
        Speaker Gender (Non-Binary)      & $0.736$    & $3.028$   & $0.003$    & $0.260$   & $1.213$   \\
        Speaker Average Age              & $-0.006$   & $-0.133$  & $0.894$    & $-0.094$  & $0.082$   \\
        \hline
        \textbf{Random Effects} & \multicolumn{5}{l}{} \\
        \hline
        Group: ResponseId (304 Levels)   & \multicolumn{5}{l}{Standard Deviation = 0.745} \\
        Group: Batch Number (10 Levels)  &  \multicolumn{5}{l}{Standard Deviation = 0.257} \\
        \hline
    \end{tabular}
    \caption{Generalized linear mixed-effects model for the identity study showing coefficients, z-values, p-values, and $95\%$ confidence intervals. 
    The model includes fixed effects for predictor variables and random intercepts by group. Predictor variables include \textit{combined duration} (continuous), \textit{listener age} (continuous), \textit{speaker average age} (continuous), as well as \textit{scriptedness} (scripted vs. unscripted), \textit{condition} (real vs. AI-clone), and \textit{gender} (categorical), with grouping by \texttt{ResponseId} (participant) and \texttt{batch\_number}.}
    \label{tab:mixedlm-results-identity}
\end{table}
\begin{table}[t]
    \centering
    \begin{tabular}{lccccc}
        \hline
        \textbf{Predictor} & \textbf{Coef.} & \textbf{z} & \textbf{p-value} & \textbf{95\% CI Lower} & \textbf{95\% CI Upper} \\ 
        \hline
        Intercept                         & $0.509$    & $5.474$   & $<0.001$ & $0.327$  & 0.691   \\
        Condition Assigned (AI generated)         & $-0.229$   & $-5.614$  & $<0.001$ & $-0.309$ & -0.149  \\
        Duration                          & $0.161$    & $6.640$    & $<0.001$ & $0.114$  & 0.209   \\
        Scripted (vs Unscripted)          & $0.311$    & $3.153$    & $0.002$    & $0.118$  & 0.505   \\
        Listener Gender (Male)            & $-0.011$    & $-0.211$   & $0.833$    & $-0.110$ & 0.089   \\
        Listener Gender (Non-Binary)      & $0.274$   & $1.233$  & $0.218$    & $-0.162$ & 0.710   \\
        Listener Age                      & $-0.000$    & $-0.162$   & $0.871$    & $-0.005$ & 0.004   \\
        Speaker Gender (Male)             & $0.020$   & $0.493$  & $0.622$    & $-0.059$ & 0.099   \\
        Speaker Gender (Non-Binary)       & $-0.126$    & $-0.866$   & $0.387$    & $-0.410$ & 0.159   \\
        Speaker Age                       & $0.028$   & $1.3590$  & $0.174$    & $-0.012$ & 0.067   \\
        \hline
        \textbf{Random Effects} & \multicolumn{5}{l}{} \\
        \hline
        Group: ResponseId (294 Levels)    & \multicolumn{5}{l}{Standard Deviation = $0.263$} \\
        Group: Batch Number (10 Levels)   & \multicolumn{5}{l}{Standard Deviation = $< 0.001$} \\
        \hline
    \end{tabular}
    \caption{Mixed-effects logistic regression model for the naturalness study showing coefficients, z-values, p-values, and $95\%$ confidence intervals. The model includes fixed effects for predictor variables and random intercepts by group. Predictor variables include \textit{duration} (continuous), \textit{listener age} (continuous), \textit{speaker age} (continuous), as well as \textit{scriptedness} (scripted vs. unscripted), \textit{condition} (real vs. fake), and \textit{gender} (categorical), with grouping by \texttt{ResponseId} (participant) and \texttt{batch\_number}.}
    \label{tab:mixedlm-results-naturalness}
\end{table}

\subsection*{Naturalness}

In the naturalness study, people were asked to directly judge whether individual audio clips were real or AI generated (irrespective of identity). This proved to be a more challenging task, Figure~\ref{fig:methodsandresults}(b). When the audio clip contained a real voice, participants were correct on average $67.4\%$ of the time (standard deviation $14.8\%$). Similarly, when the audio clip was one of our AI clones, participants were correct $60.8\%$ of the time (standard deviation $16.7\%$). Indeed, $9.7\%$ and $21.0\%$ of participants were at or below chance for the real and AI-generated stimuli, respectively. 

Nonetheless, for both stimulus types, average performance is significantly greater than chance with a medium or large effect size (real: t($299$) = $20.4$, p << $0.001$, D = $1.177$; AI generated: t($299$) = $11.2$, p << $0.001$, D = $0.648$). A pairwise comparison between stimulus types indicates a small but statistical significant difference associated with stimulus type (t($299$) = $4.6$, p << $0.001$, D = $0.263$).

Consistent with these results, the logistic regression model (Table~\ref{tab:mixedlm-results-naturalness}) indicates a significant effect of stimulus type (real vs AI generated) on accuracy. None of the listener or speaker demographics are associated with statistically significant effects. However, as in the identity study, there are, again, significant effects associated with audio duration and scriptedness. In this study, longer clips and unscripted clips are associated with more accurate judgments of naturalness.

Together with the results of the identity study, these results reflect a status quo whereby people can often be tricked into thinking that AI-generated clones have the same identity as a real speaker, and cannot reliably detect when a voice they hear is AI generated.


\section*{Discussion}

Here, we discuss the implications of these results further and report exploratory analyses of our stimuli and participant strategies.

\subsection*{Effects of Audio Duration and Scriptedness}

AI-powered phone scams can range from brief, scripted robocalls to fully-fledged conversations. Since our analyses revealed a significant effect of audio clip length and scriptedness on performance in both studies, we conducted an exploratory follow up to understand better how these factors affect performance. We focused on the naturalness task and aimed to investigate whether the effect of clip length on performance in the naturalness study was due, at least in part, to the fact that the unscripted responses tended to be longer (i.e.,~was improved performance caused by clip length or unscripted content). This study contained a new set of $25$ additional audio clips with longer scripted and shorter unscripted audio clips. It also included combined scripted-unscripted clips, for example, a speaker reading aloud a written question and then answering it. These audio clips were played to $30$ new participants. 

We first conducted exploratory analyses separately on clip length and scriptedness for this new dataset. These analyses revealed a weak to moderate (but statistically significant) positive relationship between audio duration and the accuracy of identifying the audio as real or AI generated (Spearman $r_s$ = $0.245$, p << $0.001$). While scriptedness was not associated with a statistically significant difference in performance in this study (Friedman test = $4.2$,  p < $0.125$), we observed a qualitative trend consistent with the main study, whereby the median accuracy was highest for combined audios ($83.3\%$), followed by unscripted audios ($76.7\%$) and scripted audios ($56.7\%$). 

A logistic regression that included effects for both scriptedness and audio duration indicated that these effects were not statistically significant, likely because clip length and scriptedness were still somewhat correlated even in this dataset. However, the coefficients were consistent with the findings of the logistic regression model from the naturalness study, which identified positive relationships for both duration and scripting. Therefore, these results provisionally support the conclusion that both the duration of an audio and its scriptedness influence accuracy, but a larger dataset with a wider variety of audio clips would be necessary to confirm this. 

These followup results, in conjunction with the findings from the naturalness study, suggest a potential strategy for listeners to better identify fraudulent AI-voices: engage the speaker in a longer conversation by, for example, asking open-ended questions. 

\begin{figure}[t]
    \includegraphics[width=1\textwidth]{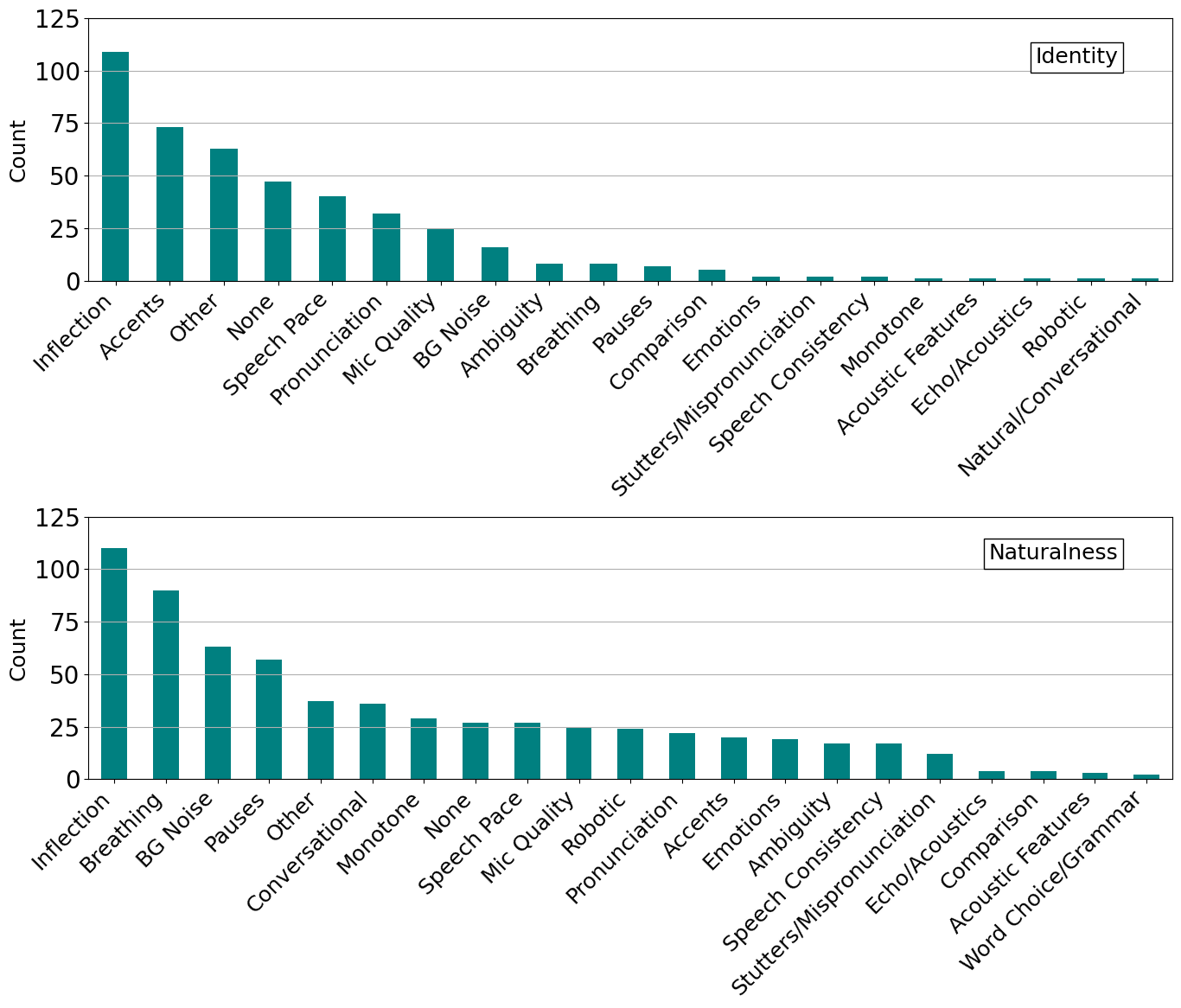}  
    \caption{The $21$ most frequent thematic codes as reported by participants in the naturalness (top) and identity (bottom) studies.}
    \label{fig:qual}
\end{figure}

\subsection*{Self-reported Listener Strategies}

At the end of each study, participants were asked to share any tactics they used to differentiate between same/different voices (identity study) and real/AI-generated voices (naturalness study). Keywords were extracted from their responses using a qualitative coding analysis and grouped into thematic codes.  As shown in Figure~\ref{fig:qual}, the top three most frequent codes in the identity study were ``inflection,'' ``breathing,'' and ``background noise.'' For the naturalness study, these were ``inflection,'' ``accent,'' and ``other.''

We also found evidence that these strategies were impacting performance by looking at participant responses. Specifically, six audio recordings were correctly classified by all participants in the naturalness study, all of which were real, and five of which were unscripted. Qualitatively, we observed that these six recordings did indeed contain audible background noise, opening mouth clicks, and several disfluencies (um, ah, etc.). At the same time, these cues were not necessarily diagnostic in general. For example, "background noise" was mentioned $79$ times across both studies, yet, upon analyzing the average background noise in both datasets, no significant effects on performance were found. Indeed, given the variety of background noise conditions that can occur on phone calls, it seems unlikely that this cue holds a key to detecting AI voice clones on the phone. We therefore suggest that a good listener strategy would be to disregard background noise as a cue and focus on the qualities of the caller's voice.

\subsection*{Gender Differences}

Although there were no effects of gender (or other demographics) in the naturalness study, we found an effect of gender in the identity study. In particular, the voices of male and non-binary speakers were associated with more ``same'' responses as compared to female speakers. This could be the result of a bias in the AI training in which female voices are underrepresented and so AI-generated male voices are more identity preserving. However, we were also curious whether this finding might reflect an additional perceptual effect related to gender. For example, the well-known cross-race effect describes the phenomena in which people more easily recognize faces that belong to our own racial group, or the group to which we have most exposure~\cite{young2012perception}. We wondered if there was a similar cross-gender effect in voice perception. 

To investigate this possibility, the main logistic regression models described in Tables~\ref{tab:mixedlm-results-identity} and~\ref{tab:mixedlm-results-naturalness} were modified to replace the speaker and listener gender predictors with a single same/different gender predictor corresponding to whether the listener and speaker were of the same gender (because non-binary listeners and speakers were heavily underrepresented in our dataset, they were excluded from this analysis). For both studies, we found no evidence for a cross-gender effect. That is, participants were not notably better or worse at detecting AI-generated voices within or across their own gender.

\subsection*{Forensic Techniques}

While modern forensic techniques~\cite{barrington2023single} are better than humans at distinguishing between certain characteristics of real and AI-generated voices, these techniques typically operate asynchronously, making it difficult to protect consumers on phone/video calls. While synchronous techniques can operate at the source of a call, this raises serious privacy concerns that would need to be addressed. Any such technology would also have to have exceedingly high accuracy to avoid false alarms and giving a false sense of security against AI-powered voice scams.

Our reporting of people's ability to detect AI-generated voices derives from a task in which their attention is fully drawn to either the identity or realism of a voice. In real-world scenarios, people may be less attentive to the voices and therefore more likely to be fooled. 

Another intervention that may prove useful (albeit not perfect) in mitigating the risk of AI-powered scams is the insertion of difficult to remove and easy to identify, imperceptible watermarks into AI-generated voices. With the appropriate software at the receiver,  AI-generated voices can be easily identified. If, however, all AI-powered voice generators do not deploy this technology, scammers would simply migrate to those services that opt out of this security protocol.

\section*{Conclusions}

The quality and realism of AI-generated media is rapidly improving. Given the results reported here, there is good reason to believe that AI-generated voices will soon be indistinguishable from real ones both in terms of naturalness and identity. While this should be considered a triumph for those on the generative side, it raises real concerns for public safety. Our results highlight that relying on human perception to detect AI-generated voice clones is no longer consistently reliable. Thus, improved technologies that can detect AI-generated voices while protecting the user's privacy will become essential tools for preventing phone-based -- and eventually, video-based -- frauds.


\section*{Data Collection and Availability}

Audio data available at \url{https://huggingface.co/datasets/faridlab/deepspeak_v1}. Anonymized speaker and listener data available at \url{https://doi.org/10.5281/zenodo.13654688}. This study was approved by the UC Berkeley Committee for Protection of Human Subjects (2023-09-16711). Participants provided informed consent prior to participation; data collection was performed in accordance with relevant guidelines and regulations.


\section*{Acknowledgments}

We thank ElevenLabs for API access to their voice cloning service. This work was supported by funding from the University of California Noyce Initiative and a gift from YouTube.

\section*{Competing interests}

HF is the co-founder and Chief Science Officer at GetReal Labs, a company developing techniques to detect AI-generated content, and he is an advisor to the Content Authentication Initiative, and is a member of the steering committee for the Coalition for Content Provenance and Authentication.

\bibliography{main}

\begin{thebibliography}{10}
\urlstyle{rm}
\expandafter\ifx\csname url\endcsname\relax
  \def\url#1{\texttt{#1}}\fi
\expandafter\ifx\csname urlprefix\endcsname\relax\def\urlprefix{URL }\fi
\expandafter\ifx\csname doiprefix\endcsname\relax\def\doiprefix{DOI: }\fi
\providecommand{\bibinfo}[2]{#2}
\providecommand{\eprint}[2][]{\url{#2}}

\bibitem{npr-biden-robocall}
\bibinfo{author}{Bond, S.}
\newblock \bibinfo{title}{A political consultant faces charges and fines for
  {B}iden deepfake robocalls}.
\newblock \bibinfo{howpublished}{NPR} (\bibinfo{year}{2024}).

\bibitem{nbc-harris-story}
\bibinfo{author}{Alsharif, M.}, \bibinfo{author}{Marquez, A.} \&
  \bibinfo{author}{Mullen, A.}
\newblock \bibinfo{title}{Elon {M}usk retweets altered {K}amala {H}arris
  campaign ad}.
\newblock \bibinfo{howpublished}{NBC News} (\bibinfo{year}{2024}).

\bibitem{aarp2024chatbots}
\bibinfo{author}{{AARP}}.
\newblock \bibinfo{journal}{\bibinfo{title}{Chatbots and voice-cloning fuel
  rise in {AI}-powered scams}}.
\newblock {\emph{\JournalTitle{AARP Arizona}}}  (\bibinfo{year}{2024}).

\bibitem{cnn-hk-story}
\bibinfo{author}{Magramo, K.} \& \bibinfo{author}{Chen, H.}
\newblock \bibinfo{title}{Finance worker pays out \$25 million after video call
  with deepfake 'chief financial officer'}.
\newblock \bibinfo{howpublished}{CNN} (\bibinfo{year}{2024}).

\bibitem{farid2022creating}
\bibinfo{author}{Farid, H.}
\newblock \bibinfo{journal}{\bibinfo{title}{Creating, using, misusing, and
  detecting deep fakes}}.
\newblock {\emph{\JournalTitle{Journal of Online Trust and Safety}}}
  \textbf{\bibinfo{volume}{1}} (\bibinfo{year}{2022}).

\bibitem{aisynthesizedfaces}
\bibinfo{author}{Nightingale, S.~J.} \& \bibinfo{author}{Farid, H.}
\newblock \bibinfo{journal}{\bibinfo{title}{{AI}-synthesized faces are
  indistinguishable from real faces and more trustworthy}}.
\newblock {\emph{\JournalTitle{Proceedings of the National Academy of
  Sciences}}} \textbf{\bibinfo{volume}{119}}, \bibinfo{pages}{e2120481119}
  (\bibinfo{year}{2022}).

\bibitem{bray2023testing}
\bibinfo{author}{Bray, S.~D.}, \bibinfo{author}{Johnson, S.~D.} \&
  \bibinfo{author}{Kleinberg, B.}
\newblock \bibinfo{journal}{\bibinfo{title}{Testing human ability to detect
  ‘deepfake’images of human faces}}.
\newblock {\emph{\JournalTitle{Journal of Cybersecurity}}}
  \textbf{\bibinfo{volume}{9}}, \bibinfo{pages}{tyad011}
  (\bibinfo{year}{2023}).

\bibitem{deepfakedetectionhumancrowd}
\bibinfo{author}{Groh, M.}, \bibinfo{author}{Epstein, Z.},
  \bibinfo{author}{Firestone, C.} \& \bibinfo{author}{Picard, R.}
\newblock \bibinfo{journal}{\bibinfo{title}{Deepfake detection by human crowds,
  machines, and machine-informed crowds}}.
\newblock {\emph{\JournalTitle{Proceedings of the National Academy of
  Sciences}}} \textbf{\bibinfo{volume}{119}}, \bibinfo{pages}{e2110013119}
  (\bibinfo{year}{2022}).

\bibitem{kobis_fooled_2021}
\bibinfo{author}{Köbis, N.~C.}, \bibinfo{author}{Doležalová, B.} \&
  \bibinfo{author}{Soraperra, I.}
\newblock \bibinfo{journal}{\bibinfo{title}{Fooled twice: {P}eople cannot
  detect deepfakes but think they can}}.
\newblock {\emph{\JournalTitle{iScience}}} \textbf{\bibinfo{volume}{24}}
  (\bibinfo{year}{2021}).

\bibitem{groh2024human}
\bibinfo{author}{Groh, M.} \emph{et~al.}
\newblock \bibinfo{journal}{\bibinfo{title}{Human detection of political speech
  deepfakes across transcripts, audio, and video}}.
\newblock {\emph{\JournalTitle{Nature communications}}}
  \textbf{\bibinfo{volume}{15}}, \bibinfo{pages}{7629} (\bibinfo{year}{2024}).

\bibitem{mai}
\bibinfo{author}{Mai, K.~T.}, \bibinfo{author}{Bray, S.},
  \bibinfo{author}{Davies, T.} \& \bibinfo{author}{Griffin, L.~D.}
\newblock \bibinfo{journal}{\bibinfo{title}{Warning: {H}umans cannot reliably
  detect speech deepfakes}}.
\newblock {\emph{\JournalTitle{PLOS ONE}}} \textbf{\bibinfo{volume}{18}},
  \bibinfo{pages}{1--20}, \url{10.1371/journal.pone.0285333}
  (\bibinfo{year}{2023}).

\bibitem{muller}
\bibinfo{author}{M\"{u}ller, N.~M.}, \bibinfo{author}{Pizzi, K.} \&
  \bibinfo{author}{Williams, J.}
\newblock \bibinfo{title}{Human perception of audio deepfakes}.
\newblock In \emph{\bibinfo{booktitle}{1st International Workshop on Deepfake
  Detection for Audio Multimedia}}, \bibinfo{pages}{85–91}
  (\bibinfo{year}{2022}).

\bibitem{traynor2024}
\bibinfo{author}{Warren, K.} \emph{et~al.}
\newblock \bibinfo{title}{Better be computer or {I}'m dumb: {A} large-scale
  evaluation of humans as audio deepfake detectors}.
\newblock In \emph{\bibinfo{booktitle}{ACM SIGSAC Conference on Computer and
  Communications Security}}, \bibinfo{pages}{2696–2710}
  (\bibinfo{year}{2024}).

\bibitem{2024deepspeak}
\bibinfo{author}{Barrington, S.}, \bibinfo{author}{Bohacek, M.} \&
  \bibinfo{author}{Farid, H.}
\newblock \bibinfo{title}{{DeepSpeak Dataset v1.0}}.
\newblock \bibinfo{howpublished}{arXiv:2408.05366} (\bibinfo{year}{2024}).

\bibitem{timit}
\bibinfo{author}{Garofolo, J.~S.} \emph{et~al.}
\newblock \bibinfo{title}{{DARPA TIMIT} acoustic phonetic continuous speech
  corpus} (\bibinfo{year}{1993}).

\bibitem{whisper}
\bibinfo{author}{Radford, A.} \emph{et~al.}
\newblock \bibinfo{title}{Robust speech recognition via large-scale weak
  supervision}.
\newblock In \emph{\bibinfo{booktitle}{40th International Conference on Machine
  Learning}}, ICML'23 (\bibinfo{year}{2023}).

\bibitem{koluguri2022titanet}
\bibinfo{author}{Koluguri, N.~R.}, \bibinfo{author}{Park, T.} \&
  \bibinfo{author}{Ginsburg, B.}
\newblock \bibinfo{title}{Tita{N}et: {N}eural model for speaker representation
  with {1D} depth-wise separable convolutions and global context}.
\newblock In \emph{\bibinfo{booktitle}{IEEE International Conference on
  Acoustics, Speech and Signal Processing}}, \bibinfo{pages}{8102--8106}
  (\bibinfo{organization}{IEEE}, \bibinfo{year}{2022}).

\bibitem{young2012perception}
\bibinfo{author}{Young, S.~G.}, \bibinfo{author}{Hugenberg, K.},
  \bibinfo{author}{Bernstein, M.~J.} \& \bibinfo{author}{Sacco, D.~F.}
\newblock \bibinfo{journal}{\bibinfo{title}{Perception and motivation in face
  recognition: {A} critical review of theories of the cross-race effect}}.
\newblock {\emph{\JournalTitle{Personality and Social Psychology Review}}}
  \textbf{\bibinfo{volume}{16}}, \bibinfo{pages}{116--142}
  (\bibinfo{year}{2012}).

\bibitem{barrington2023single}
\bibinfo{author}{Barrington, S.}, \bibinfo{author}{Barua, R.},
  \bibinfo{author}{Koorma, G.} \& \bibinfo{author}{Farid, H.}
\newblock \bibinfo{title}{Single and multi-speaker cloned voice detection:
  {F}rom perceptual to learned features}.
\newblock In \emph{\bibinfo{booktitle}{IEEE International Workshop on
  Information Forensics and Security}}, \bibinfo{pages}{1--6}
  (\bibinfo{organization}{IEEE}, \bibinfo{year}{2023}).

\end{thebibliography}

\newpage
\section*{Supplementary Information}

%
%
\begin{table}[h!]
\centering
\begin{tabular}{p{0.75\linewidth}} 
\toprule
\textbf{Naturalness (introduction and consent)} \\
\midrule
In the following survey, you will be asked to listen to 44 short audios, and answer whether you believe them to be a 'real' or a 'deepfake' (computer-generated) voice. The survey should take you about 30 minutes to complete. \\
\\
You will then be given a completion ID that you must copy in order to complete the survey task. \\
\\
Once completed, your answers will be validated. There will be 4 attention checks during the course of the study. Payment will be provided once your submission has been approved. \\
\\
Finally, your results will be anonymized and used to provide a research evaluation of how well humans can detect deepfake voices. No identifying information about you will be shared publicly. \\
\\
Please select the option below to consent to participate in this study. \\
\midrule
\textbf{Naturalness (instructions)} \\
\midrule
In this study, you will be shown \textbf{one audio at a time.} You can play the audio by clicking the 'play icon', as shown in the below screenshot: \\
\includegraphics[width=0.6\linewidth]{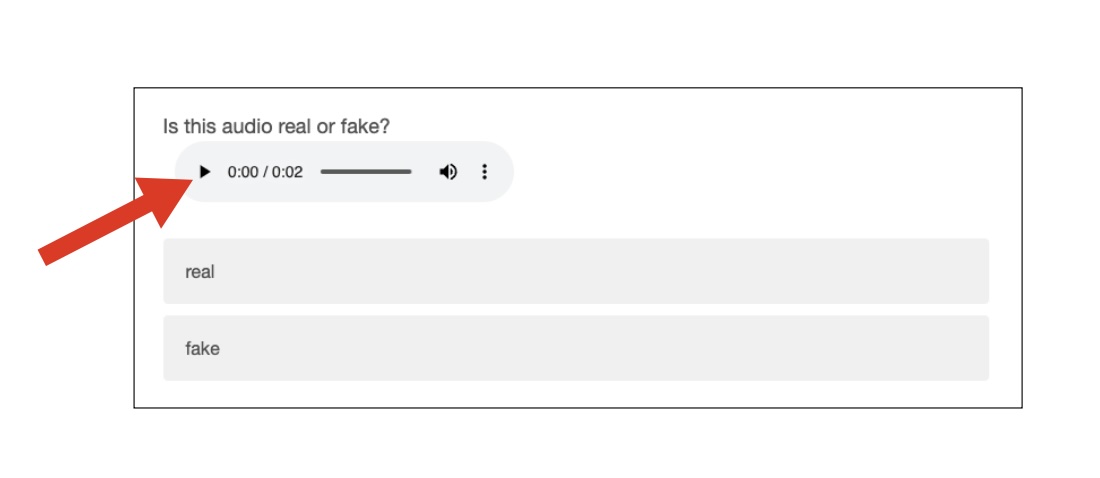} \\
A test audio is provided below. Please \textbf{play the test audio} to check that you can operate the player correctly. You can then \textbf{select the correct number} that is played in the audio. \\
\textit{audio placeholder} \\
\midrule
\textbf{Naturalness (examples)} \\
\midrule
In each question, you will be asked to select whether a voice is real or fake (computer-generated). Computer generated voices may sound very human-like, and so we appreciate that this could be difficult for you. Please try your best. \\
\\
Below are two examples of \textbf{real} voices: \\
\textit{audio placeholder} \\
\textit{audio placeholder} \\
\\
And two examples of \textbf{fake} voices: \\
\textit{audio placeholder} \\
\textit{audio placeholder} \\
\\
Please confirm that you have understood this message and have listened to the provided examples. \\
\bottomrule
\end{tabular}
\caption{Survey instructions for Naturalness study}
\end{table}
\newpage

%
%
\begin{table}[h!]
\centering
\begin{tabular}{p{0.75\linewidth}} 
\toprule
\textbf{Identity (introduction and consent)} \\
\midrule
In the following survey, you will be asked to listen to approximately 45 short audio pairs, and answer whether you believe them to be the same or different voices. The survey should take you about 45 minutes to complete. \\
\\
You will then be given a completion ID that you must copy in order to complete the survey task. \\
\\
Once completed, your answers will be validated. There will be 4 attention checks during the course of the study that must be passed in order for your submission to be considered valid. Payment will be provided once your submission has been approved. \\
\\
Finally, your results will be anonymized and used to provide a research evaluation of how well humans can identify voices. No identifying information about you will be shared publicly. \\
\\
Please select the option below to consent to participate in this study. \\
\midrule
\textbf{Identity (instructions)} \\
\midrule
In this study, you will be shown \textbf{two audios at a time.} You will be asked if you think they are spoken by the \textbf{same identity} or \textbf{different identities.} We want you to focus only on the identity of the person talking, and not how they are talking or what they are saying. \\
\\
You can play the audios by clicking the 'play' icon on each audio, as shown in the below screenshot: \\
\includegraphics[width=0.6\linewidth]{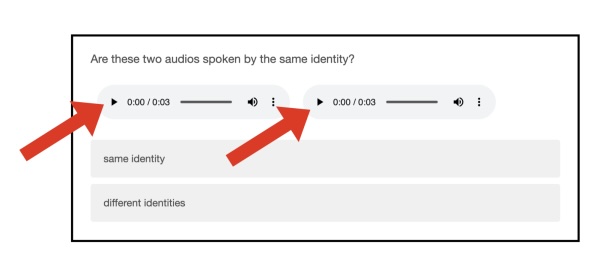} \\
Please note that you won't be able to continue to the next question until you have played both audios in full. If you experience issues with this, please refresh your browser and play both audios fully (your progress will be saved). \\
\\
A test audio is provided below. Please \textbf{play the test audio} to check that you can operate the player correctly. You can then \textbf{select the correct number} that is mentioned in the audio. \\
\textit{audio placeholder} \\
\bottomrule
\end{tabular}
\caption{Survey instructions for the Identity study}
\end{table}

\end{document}